# Progressive approximation of bound states by finite series of square-integrable functions

A. D. Alhaidari

*Saudi Center for Theoretical Physics, P.O. Box 32741, Jeddah 21438, Saudi Arabia*

**Abstract:** We use the "tridiagonal representation approach" to solve the time-independent Schrödinger equation for bound states in a basis set of finite size. We obtain two classes of solutions written as finite series of square integrable functions that support a tridiagonal matrix representation of the wave operator. The differential wave equation becomes an algebraic three-term recursion relation for the expansion coefficients of the series, which is solved in terms of finite polynomials in the energy and/or potential parameters. These orthogonal polynomials contain all physical information about the system. The basis elements in configuration space are written in terms of either the Romanovski-Bessel polynomial or the Romanovski-Jacobi polynomial. The maximum degree of both polynomials is limited by the polynomial parameter(s). This makes the size of the basis set finite but sufficient to give a very good approximation of the bound states wavefunctions that improves with an increase in the basis size.

## 1. Introduction

The Tridiagonal Representation Approach (TRA) is an algebraic method for solving linear ordinary differential equations of the second order [1-4]. It is inspired by the *J*-matrix method, which was proposed almost 45 years ago by Yamani *et al* to handle the scattering problem in quantum mechanics [5-7]. Originally, the method was considered a purely physical technique but then Ismail and his collaborators turned it into a mathematical technique [8-11]. The advantage of the method in being algebraic is reinforced by the analytic power of orthogonal polynomials and special functions. On the computational side, it is favored as being reliant on powerful numerical techniques that deal with tridiagonal matrices and Gauss quadrature. In the method, the solution of the differential equation (e.g., Schrödinger equation) is written as a convergent series (finite or infinite) in terms of a complete set of square integrable functions (basis). This set is chosen such that the matrix representation of the differential operator is tridiagonal. Consequently, the differential equation becomes a three-term recursion relation for the expansion coefficients of the series. The recursion is then solved in terms of orthogonal polynomials in the space of parameters of the differential equation (e.g., the energy and potential parameters). All physical properties of the solution (e.g., energy spectrum, scattering phase shift, density of states, etc.) are obtained from the properties of these polynomials (e.g., their weight function, generating function, zeros, asymptotics, etc.). For a detailed description of the method and how to use it for solving quantum mechanical problems, one may consult [1,12] and references therein.

Beside the tridiagonal constraint on the matrix representation of the wave operator, the basis set is required to be complete. Mathematically, completeness of a discrete basis set used for describing a continuous spectrum or an infinite discrete spectrum means that:

(i) Its size is infinite,

(i) It is defined over the whole configuration space of the problem, and

(ii) It satisfies the boundary conditions.

The last requirement usually implies that the basis elements are square integrable. Rigorously, one should also require that the basis set be dense over all regions of configuration space. For example, removing a finite subset of the infinite basis set, however small the subset is, will render it incomplete. On the other hand, a finite basis set could give an accurate description of the bound states if the system has a finite discrete spectrum. A necessary, but not sufficient condition, for a finite basis set to give a faithful physical representation of such system is that the size of the basis set must be greater than or equal to the size of the spectrum. Moreover, the finite basis could in fact give a very good approximation of bound states for a system with infinite spectrum and with an accuracy that increases with the size of the basis. Consequently, we opt to use a basis set of finite size to describe the bound states. The advantage, in this case, is that we deal with finite rather than infinite series. If we designate such basis set as $\{\phi_n(y)\}_{n=0}^{N}$ then the bound state wavefunction is written as

$$\psi(x) = \sum_{n=0}^{N} f_n \phi_n(y), \tag{1}$$

where $x$ is the coordinate of the physical configuration space, $x_- \leq x \leq x_+$, and $y$ is a dimensionless variable defined over the interval $y_- \leq y \leq y_+$ with $y_\pm = y(\lambda x_\pm)$ and $\lambda$ is scale parameter of inverse length dimension. In the TRA, elements of this finite basis set are written as $\phi_n(y) = W(y) P_n(y)$, where $W(y)$ is a positive entire function and $P_n(y)$ is a polynomial of degree $n$ in $y$ that satisfies the following conditions:

(1) It has a maximum degree $N$ whose value depends on the polynomial parameter(s).

(2) It is a solution of a second order differential equation of the form:

$$\left[ p(y)\frac{d^2}{dy^2} + q(y)\frac{d}{dy} + h(n) \right] P_n(y) = 0. \tag{2a}$$

(3) It satisfies a three-term recursion relation that reads:

$$y P_n(y) = a_n P_n(y) + b_{n-1} P_{n-1}(y) + c_n P_{n+1}(y). \tag{2b}$$

(4) Its first order differential relation (backward/forward shift operator) is as follows:

$$p(y)\frac{dP_n(y)}{dy} = d_n P_n(y) + e_{n-1} P_{n-1}(y) + w_n P_{n+1}(y). \tag{2c}$$

(5) It satisfies an orthogonality relation that reads:

$$\int_{y_-}^{y_+} \omega(y) P_n(y) P_m(y) dy = \xi_n \delta_{n,m}. \tag{2d}$$

All functions and coefficients shown above in (2a)-(2d) are well defined once a polynomial sequence $\{P_n(y)\}_{n=0}^N$ is chosen. In Appendices A and B, we list all these properties for two such polynomials that will be used in this work. Once the basis set is fixed, the bound states wavefunction (1) for a given system defined by its corresponding potential function will be fully determined if we can identify the set of expansion coefficients $\{f_n\}_{n=0}^N$. This will be done by proceeding in the TRA as follows.

Let the Schrödinger wave equation be written as $\mathcal{D}\psi(x)=0$. If we can calculate the action of the wave operator $\mathcal{D}$ on the basis element $\phi_n(y)$ then we obtain the matrix wave equation that will be solved algebraically. The TRA requires that this action take the following form [1-4, 12]

$$\begin{aligned}\mathcal{D}\phi_n(y) &= g(y)\left[u_n\phi_n(y)+s_{n-1}\phi_{n-1}(y)+t_n\phi_{n+1}(y)\right] \\ &= g(y)W(y)\left[u_nP_n(y)+s_{n-1}P_{n-1}(y)+t_nP_{n+1}(y)\right]\end{aligned} \quad (3)$$

where $\{u_n,s_n,t_n\}$ are constant coefficients and $g(y)$ is a node-less function over the open interval $y\in(y_-,y_+)$. Moreover, it is required that the integral $\langle\phi_m|\mathcal{D}|\phi_n\rangle = \lambda\int_{x_-}^{x_+}\phi_m(y)g(y)\phi_n(y)dx$ be proportional to $\delta_{m,n}$. Therefore, using the orthogonality (2d), this requirement implies that

$$g(y)W^2(y) \propto \omega(y)w(y), \quad (4)$$

where $\frac{dy}{dx}=\lambda w(y)$. This suggests that $g(y)$, $w(y)$, and thus $W(y)$, have the same functional form as the weight function $\omega(y)$. Substituting the wavefunction expansion (1) into the wave equation $\mathcal{D}\psi(x)=0$ and using the action (3) gives the following three-term recursion relation for the expansion coefficients $\{f_n\}$

$$u_nf_n+s_nf_{n+1}+t_{n-1}f_{n-1}=0. \quad (5)$$

Consequently, the solution of the wave equation $\mathcal{D}\psi(x)=0$ reduces to an algebraic solution of the discrete recursion relation (5). Now, we write $f_n=f_0F_n$, which make $F_0=1$. If we can also write $u_n=r_n-z$ where $z$ is a parameter independent of $n$ and such that $\{r_n,s_n,t_n\}$ are all independent of $z$ then this equation could be rewritten as follows

$$zF_n(z)=r_nF_n(z)+s_nF_{n+1}(z)+t_{n-1}F_{n-1}(z), \quad (6)$$

and $F_n(z)$ becomes a polynomial in $z$ of degree $n$. Moreover, if $s_nt_n>0$ for all $n$ then according to Favard theorem [13] (a.k.a. the spectral theorem; see section 2.5 in [14]) $\{F_n(z)\}$ will be a sequence of orthogonal polynomials on the real line with $f_0^2(z)$ as their positive definite weight function [1,15,16]. In the following sections, we use the TRA to obtain the bound states solutions of the time-independent Schrödinger equation in configuration space $x$ for the potential $V(x)$ and energy $E$ that reads as follows:

$$\left[-\frac{1}{2}\frac{d^2}{dx^2}+V(x)-E\right]\psi(x)=0. \quad (7)$$

where we have adopted the atomic units $\hbar = m = 1$. In 2D (or 3D) with cylindrical (or spherical) symmetry, $V(x)$ will stand for the effective potential that includes the orbital term: $V_{eff}(x) = \frac{L^2 - \frac{1}{4}}{2x^2} + V(x)$, where $L = 0, \pm 1, \pm 2, ...$ or $L = \frac{1}{2}, \frac{3}{2}, \frac{5}{2}, ...$, respectively. In what follows, we consider two classes of problems each is associated with a given finite polynomial sequence $\{P_n(y)\}_{n=0}^{N}$.

## 2. TRA in the Bessel basis

For this class of TRA solutions, we choose the basis polynomial $P_n(y)$ as the Romanovski-Bessel (R-Bessel) polynomial $Y_n^\mu(y)$ defined by (A1) in Appendix A. Therefore, the weight function associated with these polynomials as shown in the orthogonality relation (A3) suggest that we choose the coordinate transformation and basis elements as follows:

$$\frac{dy}{dx} = \lambda y^{-a} e^{-b/y}, \tag{8a}$$

$$\phi_n(y) = y^{-\alpha} e^{-\beta/y} Y_n^\mu(y), \tag{8b}$$

where the real parameters $\{a, b, \alpha, \beta, \mu\}$ are to be determined in terms of the physical parameters and energy by the TRA constraints. Moreover, $n = 0, 1, .., N$ with $N = \lfloor -\mu - \frac{1}{2} \rfloor$ where $\lfloor x \rfloor$ stands for the largest integer less than $x$. If we write the Schrödinger wave equation (7) as $\mathcal{D}\psi(x) = 0$, then for further developments we need to evaluate the action of the wave operator $\mathcal{D}$ on the basis element. The most laborious part of this calculation is the action of the kinetic operator $-\frac{1}{2}\frac{d^2}{dx^2}$ on $\phi_n(y)$. Using the differential equation (A4) for $Y_n^\mu(y)$, we obtain

$$\begin{aligned}\frac{d^2}{dx^2}\phi_n(y) = \lambda^2 y^{-(2a+\alpha+2)} e^{-(2b+\beta)/y} &\left\{ [2\beta + b - 1 - (2\alpha + a + 2\mu + 2)y] \frac{d}{dy} \right.\\ &\left. + \frac{\beta(\beta+b)}{y^2} - \frac{\beta(2\alpha+a+2)+\alpha b}{y} + \alpha(\alpha+a+1) + n(n+2\mu+1) \right\} Y_n^\mu(y)\end{aligned} \tag{9}$$

The action of the wave operator on the basis element $\phi_n(y)$ is obtained by adding the term $y^{2(a+1)} e^{2b/y} [\varepsilon - U(y)]$ inside the curly brackets, where $\varepsilon = 2E/\lambda^2$ and $U(y) = 2V(x(y))/\lambda^2$. To obtain energy independent potentials and to eliminate zero energy solutions, we must set $b = 0$. Therefore, allowed coordinate transformations must satisfy $\frac{dy}{dx} = \lambda y^{-a}$. Thus, if $a \neq -1$ then $y(x) = [(a+1)\lambda x + c]^{\frac{1}{a+1}}$ where $c$ is a constant of integration to be determined by the conditions $y(x_-) = 0$ and $y(x_+) \to \infty$. On the other hand, if $a = -1$ then $y(x) = e^{\lambda x}$ and $x_\pm = \pm\infty$. Table 1 is a list of all coordinate transformations relevant to the space of solutions in the Bessel basis. The TRA fundamental constraint (3) leads to the following three scenarios:

**Table 1**: Coordinate transformations that are compatible with the space of solutions in the Bessel basis and where $\frac{dy}{dx} = \lambda y^{-a}$.

| $a$ | $y(x)$ | $x_-$ | $x_+$ |
|---|---|---|---|
| $-\frac{1}{2}$ | $(\lambda x/2)^2$ | $0$ | $\infty$ |
| $-1$ | $e^{\lambda x}$ | $-\infty$ | $\infty$ |
| $-\frac{3}{2}$ | $(2/\lambda x)^2$ | $\infty$ | $0$ |
| $-2$ | $1/\lambda x$ | $\infty$ | $0$ |

**First TRA scenario**: $2\beta = 1$ and $2\alpha = -(2\mu + a + 2)$

In this scenario, the action of the wave operator on the basis elements becomes:

$$\mathcal{D}\phi_n(y) = -\frac{\lambda^2}{2} y^{-(2a+\alpha+2)} e^{-1/2y}$$
$$\times \left\{ \frac{1/4}{y^2} + \frac{\mu}{y} + \left(n + \mu + \tfrac{1}{2}\right)^2 - \tfrac{1}{4}(a+1)^2 + y^{2(a+1)}\left[\varepsilon - U(y)\right] \right\} Y_n^\mu(y) \quad (10)$$

Thus, $g(y) = -\tfrac{1}{2}\lambda^2 y^{-2(a+1)}$ and the requirement (4) is automatically satisfied. The three-term recursion relation (A2) for $Y_n^\mu(y)$ and the TRA fundamental constraint (3) lead to seven cases in this scenario, which are shown in Table 2. In the Table, the basis parameter $\mu = C$ and the dimensionless potential parameters $A$, $B$, and $C$ are arbitrary but constrained only by Hermiticity and the physics of the problem. For the cases with $\varepsilon = -1/4$ we must choose $\lambda^2 = -8E$ making the basis energy dependent and dictating that $E < 0$ (bound states solution).

**Second TRA scenario**: $2\beta = 1$ and $2\alpha \neq -(2\mu + a + 2)$

In this scenario, the action of the wave operator on the basis elements reads as follows:

$$\mathcal{D}\phi_n(y) = -\frac{\lambda^2}{2} y^{-(2a+\alpha+3)} e^{-1/2y} \left\{ -(2\alpha + a + 2\mu + 2) y^2 \frac{d}{dy} \right.$$
$$\left. + \frac{1/4}{y} - \left(\alpha + 1 + \tfrac{1}{2}a\right) + y\left[\alpha(\alpha + a + 1) + n(n + 2\mu + 1)\right] + y^{2a+3}\left[\varepsilon - U(y)\right] \right\} Y_n^\mu(y) \quad (11)$$

Thus, $g(y) = -\tfrac{1}{2}\lambda^2 y^{-(2a+3)}$ and the requirement (4) dictates that $2\alpha = -(2\mu + a + 3)$. The differential relation (A8) for $Y_n^\mu(y)$ and its three-term recursion relation (A2) show that the TRA fundamental constraint (3) leads to three cases that are physically equivalent to those at the bottom three rows of Table 2.

**Third TRA scenario**: $2\beta \neq 1$ and $2\alpha = -(2\mu + a + 2)$

In this scenario, the action of the wave operator on the basis elements reads as follows:

$$\mathcal{D}\phi_n(y) = -\frac{\lambda^2}{2} y^{-(2a+\alpha+4)} e^{-\beta/y} \left\{ (2\beta - 1) y^2 \frac{d}{dy} \right.$$

$$\left. + \beta^2 + 2\beta\mu y + y^2 \left[ \left(n + \mu + \tfrac{1}{2}\right)^2 - \tfrac{1}{4}(a+1)^2 \right] + y^{2(a+2)}[\varepsilon - U(y)] \right\} Y_n^\mu(y) \tag{12}$$

Thus, $g(y) = -\tfrac{1}{2}\lambda^2 y^{-2(a+2)}$ and the requirement (4) dictates that $2\alpha = -(2\mu + a + 4)$ and $2\beta = 1$, which is in conflict with the assumptions made for this scenario. Therefore, this scenario will be ignored.

In the following section, we give an illustrative example where we consider a physically interesting system that corresponds to one of the cases from the first scenario above and obtain the corresponding bound states solutions.

**Table 2:** The class of potential functions that belong to the solution space in the Bessel basis. The potential parameters $A$, $B$, and $C$ are arbitrary with the basis parameter $\mu = C$.

| $a$ | $y(x)$ | $-2g(y)/\lambda^2$ | $2V(x)/\lambda^2$ | $2E/\lambda^2$ |
|---|---|---|---|---|
| $-\tfrac{1}{2}$ | $(\lambda x/2)^2$ | $y^{-1}$ | $\dfrac{1/4}{y^3} + \dfrac{C}{y^2} + \dfrac{B}{y}$ | --- |
| $-1$ | $e^{\lambda x}$ | $1$ | $\dfrac{1/4}{y^2} + \dfrac{C}{y} + Ay$ | --- |
| $-\tfrac{3}{2}$ | $(2/\lambda x)^2$ | $y$ | $\dfrac{1/4}{y} + Ay^2 + By$ | $-\mu$ |
| $-2$ | $1/\lambda x$ | $y^2$ | $Ay^3 + By^2 + Cy$ | $-1/4$ |
| $-1$ | $e^{\lambda x}$ | $y^{-1}$ | $\dfrac{1/4}{y^2} + \dfrac{B}{y}$ | --- |
| $-\tfrac{3}{2}$ | $(2/\lambda x)^2$ | $1$ | $\dfrac{1/4}{y} + Ay$ | --- |
| $-2$ | $1/\lambda x$ | $y$ | $Ay^2 + By$ | $-1/4$ |

## 3. Sample TRA solution in the Bessel basis

In spherical coordinates, the potential function associated with a local electric charge distribution as seen by an electron at distances much larger than the size of the distribution becomes a multipole expansion, which when written up to and including the linear electric quadrupole, reads as follows (see, for example, Ref. [17])

$$V(\vec{r}) = -\frac{Z}{r} - d\frac{\cos\theta}{r^2} + q\frac{\frac{1}{2}(3\cos^2\theta - 1)}{r^3}, \tag{13}$$

where $Z$ is the net positive electric charge, $d$ is the electric dipole moment along the positive z-axis, and $q$ is the linear electric quadrupole moment of the charge distribution. We took the Bohr radius, $a_0 = 4\pi\varepsilon_0\hbar^2/me^2 = 4\pi\varepsilon_0/e^2$, as the unit of length with $m$ and $-e$ being the mass and charge of the electron. Now, the quadrupole term in (13) destroys separability of the wave equation. Therefore, we consider an effective electric quadrupole interaction where the angular factor $\frac{1}{2}(3\cos^2\theta - 1)$ is replaced by a dimensionless angular parameter $\eta$ such that $-\frac{1}{2} \leq \eta \leq +1$ since $0 \leq \cos^2\theta \leq 1$. This results in an effective quadrupole potential $Q/r^3$, where the effective electric quadrupole moment is $Q = \eta q$. Consequently, the radial part of the wave equation in the atomic units becomes

$$\left[ -\frac{1}{2}\frac{d^2}{dr^2} + \frac{\gamma(\gamma+1)}{2r^2} - \frac{Z}{r} + \frac{Q}{r^3} - E \right]\psi(r) = 0, \tag{14}$$

where $\gamma$ is a quantum number that depends on the electric dipole moment $d$ and the azimuthal angular momentum quantum number $m = 0, \pm 1, \pm 2,..$ [18]. If $d = 0$ then $\gamma$ becomes the orbital angular momentum quantum number $\ell = 0, 1, 2, ...$. On the other hand, for a non-zero dipole moment $d$ and azimuthal quantum number $m$, Equations (3.6) and (3.8) in Ref. [18] give the value of $\left(\gamma + \frac{1}{2}\right)^2$ as one of the positive eigenvalues of an infinite symmetric tridiagonal matrix whose elements are

$$T_{i,j} = \left(i + m + \frac{1}{2}\right)^2 \delta_{i,j} - d\sqrt{\frac{i(i+2m)}{(i+m)^2 - 1/4}}\,\delta_{i,j+1} - d\sqrt{\frac{(i+1)(i+2m+1)}{(i+m+1)^2 - 1/4}}\,\delta_{i,j-1}. \tag{15}$$

Comparing the radial wave equation (14) to Eq. (7) with $x = r \geq 0$, we find that the potential function

$$V(r) = \frac{\gamma(\gamma+1)}{2r^2} - \frac{Z}{r} + \frac{Q}{r^3}, \tag{16}$$

corresponds to the fourth row of Table 2 with $a = -2$, $y(r) = 1/\lambda r$, $\lambda = 2\sqrt{-2E}$, and

$$A = 4Q\sqrt{-2E}, \quad B = \gamma(\gamma+1), \quad C = \mu = -Z/\sqrt{-2E}. \tag{17}$$

Moreover, the basis element becomes $\phi_n(y) = (\lambda r)^{-\mu} e^{-\lambda r/2} Y_n^{\mu}(1/\lambda r)$. Now, since $\mu$ is negative then the electric charge $Z$ must be positive. Moreover, since the size of the basis is $N+1$ with $N = \lfloor -\mu - \frac{1}{2} \rfloor$ then for a given Z, the basis size increases as $|E|$ decreases. This implies that the accuracy of our solution, which increases with $N$, improves as the energy level increases (i.e., for smaller $|E|$). This is in contrast with common experience that for such an electronic system the lowest energy levels are the most accurate with the ground state being the utmost accurate.

This observation will be confirmed by our findings below. Substituting $y^{2(a+1)}\left[\varepsilon - U(y)\right]$ with $a = -2$ and $\varepsilon = -1/4$ into (10) using (16), we obtain

$$\mathcal{D}\phi_n(y) = -4AEy^{\mu+2}e^{-1/2y}\left[-\frac{1}{A}\left(n+\mu+\tfrac{1}{2}\right)^2 + \frac{1}{A}\left(\gamma+\tfrac{1}{2}\right)^2 + y\right]Y_n^\mu(y). \tag{18}$$

Using the three-term recursion relation (A2) for $Y_n^\mu(y)$ we obtain the TRA fundamental equation (3) with $g(y) = -AEy^2$ and

$$r_n = \frac{-2\mu}{(n+\mu)(n+\mu+1)} - \frac{4}{A}\left(n+\mu+\tfrac{1}{2}\right)^2, \qquad z = -\frac{1}{A}(2\gamma+1)^2, \tag{19a}$$

$$t_n = \frac{n+2\mu+1}{(n+\mu+1)\left(n+\mu+\tfrac{1}{2}\right)}, \qquad s_{n-1} = \frac{-n}{(n+\mu)\left(n+\mu+\tfrac{1}{2}\right)}. \tag{19b}$$

The three-term recursion relation (6) with these coefficients and argument $z$ is identical to the recursion relation satisfied by the orthogonal polynomial $B_n^\mu(z;\gamma)$ shown in Appendix A by Eq. (A10) if only we can make the exchange $s_n \leftrightarrow t_n$. Thus, if we define $G_n = \prod_{m=0}^{n-1}(s_m/t_m)$ with $G_0 = 1$ then $F_n(z) = G_n B_n^\mu(z;-4/A)$. Finally, the $k$th bound state wavefunction (1) reads

$$\psi_k(r) = f_0(z_k)(\lambda_k r)^{-\mu_k} e^{-\lambda_k r/2} \sum_{n=0}^{N} G_n B_n^{\mu_k}(z_k;-4/A_k) Y_n^{\mu_k}(1/\lambda_k r), \tag{20a}$$

Where

$$\lambda_k = 2\sqrt{-2E_k}, \quad \mu_k = -2Z/\lambda_k, \quad A_k = 2Q\lambda_k, \quad z_k = -(2\gamma+1)^2/A_k. \tag{20b}$$

The energy spectrum $\{E_k\}_{k=0}^{N}$ is obtained using $z$ in (19a) and the spectrum formula for $B_n^\mu(z;\gamma)$ that gives $\{z_k\}_{k=0}^{N}$. Unfortunately, the analytic properties of this polynomials (e.g., weight function, generating function, asymptotics, zeros, etc.) are not yet known. This is an open problem in orthogonal polynomials along with other similar ones. For a discussion about these open problems, one may consult References [19-21]. Table 3 is a numerical evaluation of the lower part of the energy spectrum for a given set of physical parameters $\{Z,Q,\gamma\}$ associated with a given electric dipole $d$, quadrupole $q$, azimuthal quantum number $m$, and angular parameter $\eta$. In Figure 1, the red solid curve is the wavefunctions (20) corresponding to the lowest bound states energies in Table 3. We superimpose on those (as a blue dotted trace) the wavefunctions calculated using a robust numerical routine. The figure shows that the match between the two results improves as the energy levels increase, which confirms the observation made below Eq. (17).

**Table 3:** The lowest part of the binding energy (in atomic units) of the electron to the electro-static charge distribution problem discussed in section 3. The physical parameters are taken as: $Z = 2$, $d = 3$, $q = 5$, and $m = 0$. We also took $\left(\gamma+\tfrac{1}{2}\right)^2$ as the lowest positive eigenvalue of the tridiagonal symmetric matrix (15). The angular parameter was taken as $\eta = \frac{1}{2}$ making the effective quadrupole moment $Q = 2.5$.

| k | $-E_k$ |
|---|---|
| 0 | 0.278 416 771 |
| 1 | 0.148 519 404 |
| 2 | 0.091 908 602 |
| 3 | 0.062 380 108 |
| 4 | 0.045 079 654 |
| 5 | 0.034 087 447 |
| 6 | 0.026 673 470 |
| 7 | 0.021 438 481 |
| 8 | 0.017 605 589 |
| 9 | 0.014 715 516 |

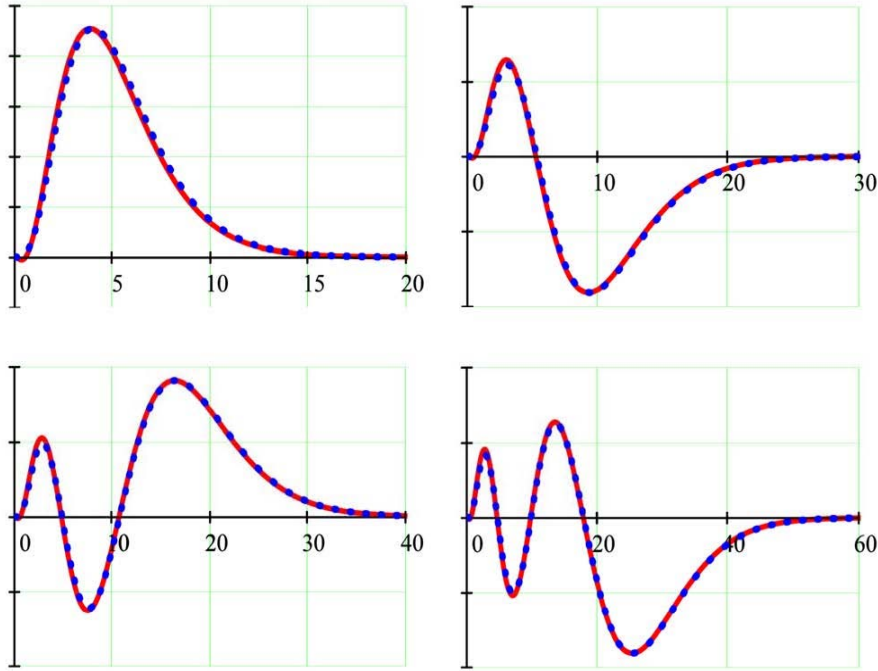

**Fig. 1:** The lowest four electron bound states in the electrostatic multipole problem, which is given by Eq. (20) and shown as a red solid curve. Superimposed on those, with blue dotted trace, are the wavefunctions calculated using a robust numerical routine. The figure shows that the match between the two results improves as the energy level increases. The problem parameters are given in the caption of Table 3 and the horizontal axis is the radial coordinate in units of the Bohr radius.

## 4. TRA in the Jacobi basis

For this class of TRA solutions, we choose the basis polynomial $P_n(y)$ as the finite Romanovski-Jacobi (R-Jacobi) polynomial $J_n^{(\mu,\nu)}(y)$ defined on the semi-infinite real line $y \geq 1$ as shown in Appendix B. The weight function associated with this polynomial as shown in the orthogonality relation (B5) or (B6) suggests that we choose the coordinate transformation and basis elements as follows:

$$\frac{dy}{dx} = \lambda(y-1)^a(y+1)^{-b}, \tag{21a}$$

$$\phi_n(y) = (y-1)^\alpha (y+1)^{-\beta} J_n^{(\mu,\nu)}(y), \tag{21b}$$

where the real parameters $\{a,b,\alpha,\beta,\mu,\nu\}$ are to be determined in terms of the physical parameters and energy by the TRA constraints. Moreover, $n = 0,1,..,N$ with $N = \left\lfloor -\frac{\mu+\nu+1}{2} \right\rfloor$. Using the differential equation (B2) for $J_n^{(\mu,\nu)}(y)$, we obtain the action of $\frac{d^2}{dx^2}$ on the basis element (21b) as follows:

$$\frac{d^2}{dx^2}\phi_n(y) = \lambda^2(y-1)^{2a+\alpha-1}(y+1)^{-2b-\beta-1} \times$$

$$\left\{ \left[(2\alpha+a-\mu-1)(y+1) - (2\beta+b+\nu+1)(y-1)\right]\frac{d}{dy} \right. \tag{22}$$

$$\left. + \frac{2\alpha(\alpha+a-1)}{y-1} - \frac{2\beta(\beta+b+1)}{y+1} + \left[(\alpha-\beta)(\alpha-\beta+a-b-1) + n(n+\mu+\nu+1)\right] \right\} J_n^{(\mu,\nu)}(y)$$

The action of the wave operator on the basis element $\phi_n(y)$ is obtained by adding the term $(y-1)^{1-2a}(y+1)^{1+2b}[\varepsilon - U(y)]$ inside the curly brackets. Table 4 is a list of all coordinate transformations relevant to this Jacobi basis. The TRA fundamental constraint (3) leads to the following three scenarios:

**First TRA scenario**: $2\alpha = \mu - a + 1$ and $2\beta = -(\nu+b+1)$

In this scenario, the action of the wave operator on the basis elements reads as follows:

$$\mathcal{D}\phi_n(y) = -\frac{\lambda^2}{2}(y-1)^{2a+\alpha-1}(y+1)^{-2b-\beta-1}\left\{(y-1)^{1-2a}(y+1)^{1+2b}[\varepsilon - U(y)]\right.$$

$$\left. + \frac{1}{2}\frac{\mu^2-(a-1)^2}{y-1} - \frac{1}{2}\frac{\nu^2-(b+1)^2}{y+1} + \frac{1}{4}\left[(2n+\mu+\nu+1)^2-(a-b-1)^2\right]\right\} J_n^{(\mu,\nu)}(y) \tag{23}$$

Thus, $g(y) = -\frac{1}{2}\lambda^2(y-1)^{2a-1}(y+1)^{-2b-1}$ and the requirement (4) is automatically satisfied. The three-term recursion relation (B3) for $J_n^{(\mu,\nu)}(y)$, the energy independence of the potential, and the TRA fundamental constraint (3) lead to the sixteen cases shown in Table 5 and Table 6. In these tables, $\mu^2 = C + (a-1)^2$, $\nu^2 = D + (b+1)^2$ and the dimensionless potential parameters $A$, $B$, $C$ and $D$ are arbitrary but constrained only by Hermiticity and the physics of the problem.

For example, $C \geq -(a-1)^2$ and $D \geq -(b+1)^2$. It should also be noted that the five cases in the bottom five rows of Table 6 are obtained from those in the top five rows by the map:

$$y \mapsto -y, \quad a \leftrightarrow -b, \quad A \leftrightarrow B, \quad C \leftrightarrow D, \quad \alpha \leftrightarrow -\beta, \quad \mu \leftrightarrow -\nu, \tag{24a}$$

$$E \leftrightarrow (-1)^{2(a+b)} E, \quad V \leftrightarrow (-1)^{2(a+b)} V. \tag{24b}$$

**Table 4:** Coordinate transformations that are compatible with the space of solutions in the Jacobi basis and where $\frac{dy}{dx} = \lambda(y-1)^a(y+1)^{-b}$.

| $(a,b)$ | $y(x)$ | $x_-$ | $x_+$ | $\bar{\lambda}$ |
|---|---|---|---|---|
| $(\frac{1}{2},-\frac{1}{2})$ | $\cosh(\lambda x)$ | 0 | $\infty$ | $\lambda$ |
| $(\frac{1}{2},0)$ | $(\bar{\lambda}x)^2 + 1$ | 0 | $\infty$ | $\lambda/2$ |
| $(0,-\frac{1}{2})$ | $2(1+\bar{\lambda}x)^2 - 1$ | 0 | $\infty$ | $\lambda/2\sqrt{2}$ |
| $(1,-\frac{1}{2})$ | $\frac{2}{\tanh^2(-\bar{\lambda}x)} - 1$ | $\pm\infty$ | 0 | $\lambda/\sqrt{2}$ |
| $(\frac{1}{2},-1)$ | $2\tan^2(\bar{\lambda}x) + 1$ | 0 | $\pi/2\bar{\lambda}$ | $\lambda/\sqrt{2}$ |
| $(1,0)$ | $e^{\lambda x} + 1$ | $-\infty$ | $\infty$ | $\lambda$ |
| $(0,-1)$ | $2e^{\lambda x} - 1$ | 0 | $\infty$ | $\lambda$ |
| $(1,-1)$ | $-1/\tanh(\lambda x)$ | $-\infty$ | 0 | $\lambda$ |

**Second TRA scenario**: $2\alpha = \mu - a + 1$ and $2\beta \neq -(\nu + b + 1)$

In this scenario, the action of the wave operator on the basis elements becomes:

$$\mathcal{D}\phi_n(y) = -\frac{\lambda^2}{2}(y-1)^{2a+\alpha-1}(y+1)^{-2b-\beta-2} \times$$

$$\left\{ -(2\beta+b+\nu+1)(y^2-1)\frac{d}{dy} + (y-1)^{1-2a}(y+1)^{2+2b}[\varepsilon - U(y)] - 2\beta(\beta+b+1) \right. \tag{25}$$

$$\left. + \frac{\mu^2 - (a-1)^2}{2}\left(1 + \frac{2}{y-1}\right) + (y+1)\big[(\alpha-\beta)(\alpha-\beta+a-b-1) + n(n+\mu+\nu+1)\big] \right\} J_n^{(\mu,\nu)}$$

Thus, $g(y) = -\frac{1}{2}\lambda^2 (y-1)^{2a-1}(y+1)^{-2b-2}$ and the requirement (4) dictates that $2\beta = -(v+b+2)$. The differential relation (B4) for $J_n^{(\mu,v)}(y)$, its three-term recursion relation (B3), and energy independence of the potential show that the TRA fundamental constraint (3) leads to five cases that are physically equivalent to those at the bottom five rows of Table 6.

**Table 5:** The class of potential functions that belong to the solution space in the Jacobi basis with $g(y) = -\frac{1}{2}\lambda^2 (y-1)^{2a-1}(y+1)^{-2b-1}$. The potential parameters $A$, $B$, $C$ and $D$ are arbitrary with the basis parameter $\mu^2 = C + (a-1)^2$ and $v^2 = D + (b+1)^2$.

| $(a,b)$ | $y(x)$ | $-2g(y)/\lambda^2$ | $2V(x)/\lambda^2$ | $2E/\lambda^2$ |
|---|---|---|---|---|
| $(\frac{1}{2},-\frac{1}{2})$ | $\cosh(\lambda x)$ | $1$ | $\frac{C/2}{y-1} - \frac{D/2}{y+1} + A(y-1) + B(y+1)$ | --- |
| $(\frac{1}{2},0)$ | $(\bar\lambda x)^2 + 1$ | $\frac{1}{y+1}$ | $\frac{C/2}{y^2-1} - \frac{D/2}{(y+1)^2} + A\frac{y-1}{y+1}$ | --- |
| $(0,-\frac{1}{2})$ | $2(1+\bar\lambda x)^2 - 1$ | $\frac{1}{y-1}$ | $\frac{C/2}{(y-1)^2} - \frac{D/2}{y^2-1} + B\frac{y+1}{y-1}$ | --- |
| $(1,-\frac{1}{2})$ | $\frac{2}{\tanh^2(-\bar\lambda x)} - 1$ | $y-1$ | $\frac{D}{y+1} + A(y-1)^2 + B(y^2-1)$ | $-\frac{\mu^2}{2}$ |
| $(\frac{1}{2},-1)$ | $2\tan^2(\bar\lambda x) + 1$ | $y+1$ | $\frac{C}{y-1} + A(y^2-1) + B(y+1)^2$ | $\frac{v^2}{2}$ |
| $(1,-1)$ | $-1/\tanh(\lambda x)$ | $y^2-1$ | $\frac{C-D}{2}y + (y^2-1)[A(y-1) + B(y+1)]$ | $\frac{\mu^2 + v^2}{-2}$ |

**Third TRA scenario:** $2\alpha \ne \mu - a + 1$ and $2\beta = -(v+b+1)$

In this scenario, the action of the wave operator on the basis elements becomes:

$$\mathcal{D}\phi_n(y) = -\frac{\lambda^2}{2}(y-1)^{2a+\alpha-2}(y+1)^{-2b-\beta-1} \times$$

$$\left\{ (2\alpha + a - \mu - 1)(y^2-1)\frac{d}{dy} + (y-1)^{2-2a}(y+1)^{1+2b}[\varepsilon - U(y)] + 2\alpha(\alpha + a - 1) \right. \quad (26)$$

$$\left. + \frac{v^2 - (b+1)^2}{2}\left(\frac{2}{y+1} - 1\right) + (y-1)[(\alpha-\beta)(\alpha-\beta+a-b-1) + n(n+\mu+v+1)] \right\} J_n^{(\mu,v)}$$

Thus, $g(y) = -\frac{1}{2}\lambda^2 (y-1)^{2a-2}(y+1)^{-2b-1}$ and the requirement (4) dictates that $2\alpha = \mu - a + 2$. The differential relation (B4) for $J_n^{(\mu,v)}(y)$, its three-term recursion relation (B3), and energy independence of the potential show that the TRA fundamental constraint (3) leads to five cases that are physically equivalent to those at the top five rows of Table 6.

**Table 6:** The class of potential functions that belong to the solution space in the Jacobi basis. The cases in the top five rows correspond to $g(y) = -\frac{1}{2}\lambda^2 (y-1)^{2a-2} (y+1)^{-2b-1}$ with $v^2 = D + (b+1)^2$. The cases in the bottom five rows correspond to $g(y) = -\frac{1}{2}\lambda^2 (y-1)^{2a-1}(y+1)^{-2b-2}$ with $\mu^2 = C + (a-1)^2$.

| $(a,b)$ | $y(x)$ | $-2g(y)/\lambda^2$ | $2V(x)/\lambda^2$ | $2E/\lambda^2$ |
|---|---|---|---|---|
| $(1,-\frac{1}{2})$ | $\dfrac{2}{\tanh^2(-\bar{\lambda}x)} - 1$ | $1$ | $\dfrac{D}{y+1} + A(y-1) + B(y+1)$ | --- |
| $(\frac{1}{2},-\frac{1}{2})$ | $\cosh(\lambda x)$ | $\dfrac{1}{y-1}$ | $\dfrac{D}{y^2-1} + B\dfrac{y+1}{y-1}$ | --- |
| $(1,0)$ | $e^{\lambda x}+1$ | $\dfrac{1}{y+1}$ | $\dfrac{D}{(y+1)^2} + A\dfrac{y-1}{y+1}$ | --- |
| $(1,-1)$ | $-1/\tanh(\lambda x)$ | $y+1$ | $A(y^2-1) + B(y+1)^2$ | $-v^2$ |
| $(\frac{1}{2},-1)$ | $2\tan^2(\bar{\lambda}x)+1$ | $\dfrac{y+1}{y-1}$ | $A(y+1) + B\dfrac{(y+1)^2}{y-1}$ | $\dfrac{v^2}{2}$ |
| $(\frac{1}{2},-1)$ | $2\tan^2(\bar{\lambda}x)+1$ | $1$ | $\dfrac{C}{y-1} + A(y-1) + B(y+1)$ | --- |
| $(\frac{1}{2},-\frac{1}{2})$ | $\cosh(\lambda x)$ | $\dfrac{1}{y+1}$ | $\dfrac{C}{y^2-1} + A\dfrac{y-1}{y+1}$ | --- |
| $(0,-1)$ | $2e^{\lambda x}-1$ | $\dfrac{1}{y-1}$ | $\dfrac{C}{(y-1)^2} + B\dfrac{y+1}{y-1}$ | --- |
| $(1,-1)$ | $-1/\tanh(\lambda x)$ | $y-1$ | $A(y-1)^2 + B(y^2-1)$ | $-\mu^2$ |
| $(1,-\frac{1}{2})$ | $\dfrac{2}{\tanh^2(-\bar{\lambda}x)} - 1$ | $\dfrac{y-1}{y+1}$ | $A\dfrac{(y-1)^2}{y+1} + B(y-1)$ | $-\dfrac{\mu^2}{2}$ |

In the following section, we give two illustrative examples. In the first, we consider a novel physical system that corresponds to one of the cases from the first scenario shown in Table 5 and obtain the bound states wavefunctions. The second example, which corresponds to one of the cases in Table 6, is an exactly solvable problem chosen for the purpose of comparison and to demonstrate the viability of our approach.

## 5. Sample TRA solutions in the Jacobi basis

We start by considering a novel system associated with the following potential model

$$V(x) = \frac{\rho^2}{(\rho x)^2 + 2}\left[\frac{C}{(\rho x)^2} - \frac{D}{(\rho x)^2 + 2} - 4A\right], \tag{27}$$

where $x \geq 0$, $C > -\frac{1}{4}$, $D > -1$ and $\rho$ is a positive scale parameter of inverse length dimension. This potential is inverse-squared singular at the origin with a singularity strength $C/2$ and it vanishes at infinity. It does not belong to the known class of exactly solvable potentials but it has a very rich structure. If $D > C - 4A$, then the potential will have two local extrema (a minimum and a maximum) and the associated energy spectrum will then contain a mixture of bound states and resonances. On the other hand, if $A > 0$ then the potential will have only one local minimum and the associated energy spectrum contains bound states with no resonances. Here, we will only be concerned with this case where $A > 0$. If we choose the dimensionless variable $y(x) = (\rho x)^2 + 1$, then the potential (27) corresponds to the case shown in the second row of Table 5 with $(a,b) = (\frac{1}{2}, 0)$ and $\lambda = 2\rho$. Moreover, the basis element becomes $\phi_n(y) = (y-1)^\alpha (y+1)^{-\beta} J_n^{(\mu,\nu)}(y)$ with $\mu = \sqrt{C + \frac{1}{4}}$, $\nu = -\sqrt{D+1}$, $2\alpha = \mu + \frac{1}{2}$, and $2\beta = -(\nu+1)$. The maximum size of the basis set, which is $N+1$, becomes $\left\lfloor \frac{1}{2}\left(\sqrt{D+1} - \sqrt{C+\frac{1}{4}} - 1\right) \right\rfloor + 1$. Substituting $(y-1)^{1-2a}(y+1)^{1+2b}[\varepsilon - U(y)]$ with $a = \frac{1}{2}$ and $b = 0$ into (23) using (27), we obtain

$$\mathcal{D}\phi_n(y) = -E(y-1)^\alpha (y+1)^{-\beta-1}\left\{\frac{1}{\varepsilon}\left[\left(n + \frac{\mu+\nu+1}{2}\right)^2 - \frac{1}{16}\right] + \frac{2A}{\varepsilon} + 1 + y\right\} J_n^{(\mu,\nu)}(y). \tag{28}$$

where $\varepsilon = E/2\rho^2$. Using the three-term recursion relation (B3) for $J_n^{(\mu,\nu)}(y)$ we obtain the TRA fundamental equation (3) with $g(y) = -E/(y+1)$ and

$$r_n = \frac{\nu^2 - \mu^2}{(2n+\mu+\nu)(2n+\mu+\nu+2)} + \frac{1}{\varepsilon}\left[\left(n + \frac{\mu+\nu+1}{2}\right)^2 - \frac{1}{16}\right], \quad z = -(2A/\varepsilon) - 1, \tag{29a}$$

$$t_n = \frac{2(n+1)(n+\mu+\nu+1)}{(2n+\mu+\nu+1)(2n+\mu+\nu+2)}, \qquad s_{n-1} = \frac{2(n+\mu)(n+\nu)}{(2n+\mu+\nu)(2n+\mu+\nu+1)}, \tag{29b}$$

The three-term recursion relation (6) with these coefficients and argument $z$ is identical to the recursion relation satisfied by the orthogonal polynomial $H_n^{(\mu,\nu)}(\zeta;\tau,\theta)$ shown in Appendix B by Eq. (B7) if only we can make the exchange $s_n \leftrightarrow t_n$. Thus, if we define $G_n = \prod_{m=0}^{n-1}(s_m/t_m)$ with $G_0 = 1$ and $E < -2A\rho^2$ then $F_n(z) = G_n H_n^{(\mu,\nu)}(\zeta;\tau,\theta)$ with

$$\tau = \frac{1}{4}, \qquad \zeta^2 = \frac{-1/4}{A(A+\varepsilon)}, \qquad \cos\theta = -(2A/\varepsilon) - 1. \tag{30}$$

The *k*th bound state wavefunction (1) reads

–14–

$$\psi_k(x) = f_0(z_k)(\rho x)^{\mu+\frac{1}{2}}\left[(\rho x)^2 + 2\right]^{\frac{\nu+1}{2}} \sum_{n=0}^{N} G_n H_n^{(\mu,\nu)}\left(\zeta_k; \tfrac{1}{4}, \theta_k\right) J_n^{(\mu,\nu)}(y), \quad (31)$$

where $z_k = \cos\theta_k = -(2A/\varepsilon_k) - 1$ and $\zeta_k^2 = -1/4A(A+\varepsilon_k)$. However, if $0 > E > -2A\rho^2$ then $F_n(z) = G_n \tilde{H}_n^{(\mu,\nu)}(\zeta;\tau,\theta)$ with $\zeta^2 = 1/4A(A+\varepsilon)$, $\cosh\theta = -(2A/\varepsilon) - 1$, and

$$\psi_k(x) = f_0(z_k)(\rho x)^{\mu+\frac{1}{2}}\left[(\rho x)^2 + 2\right]^{\frac{\nu+1}{2}} \sum_{n=0}^{N} G_n \tilde{H}_n^{(\mu,\nu)}\left(\zeta_k; \tfrac{1}{4}, \theta_k\right) J_n^{(\mu,\nu)}(y). \quad (32)$$

The second and final example is the trigonometric Pöschl-Teller potential, which could be written as follows

$$V(x) = \frac{V_0}{\sin^2(\rho x)} + \frac{V_1}{\cos^2(\rho x)}, \quad (33)$$

where $0 \leq x \leq \pi/2\rho$ and $V_i > -\rho^2/8$. If we choose the coordinate transformation $y(x) = 2\tan^2(\rho x) + 1$, then this potential corresponds to the case shown in the fifth row of Table 6 with $(a,b) = (\tfrac{1}{2}, -1)$, $\lambda = \rho\sqrt{2}$, and

$$A = \frac{V_0 + V_1}{2\rho^2}, \qquad C = -4B = 2V_0/\rho^2. \quad (34)$$

The basis element becomes $\phi_n(y) = (y-1)^\alpha (y+1)^{-\beta} J_n^{(\mu,\nu)}(y)$ with $2\alpha = \mu + \tfrac{1}{2}$, $2\beta = -\nu$, $\mu = \sqrt{\tfrac{1}{4} + 2V_0/\rho^2}$, and $\nu$ is negative but arbitrary constrained only by $\nu < -2N - \mu - 1$. Substituting $(y-1)^{1-2a}(y+1)^{1+2b}[\varepsilon - U(y)]$ with $a = \tfrac{1}{2}$ and $b = -1$ into (23) using (33), we obtain

$$\mathcal{D}\phi_n(y) = -\rho^2(y-1)^\alpha(y+1)^{-\beta}\left\{\frac{E}{\rho^2} - \frac{\nu^2}{2} + (y+1)\left[\left(n + \tfrac{\mu+\nu+1}{2}\right)^2 - \frac{\sigma^2}{4}\right]\right\} J_n^{(\mu,\nu)}(y), \quad (35)$$

where $\sigma = \sqrt{\tfrac{1}{4} + 2V_1/\rho^2}$. Using the three-term recursion relation (B3) for $J_n^{(\mu,\nu)}(y)$ we obtain the TRA fundamental equation (3) with $g(y) = -\rho^2$, $z = -E/\rho^2$, and

$$r_n = \left\{\frac{\nu^2 - \mu^2}{(2n+\mu+\nu)(2n+\mu+\nu+2)} + 1\right\}\left[\left(n + \tfrac{\mu+\nu+1}{2}\right)^2 - \frac{\sigma^2}{4}\right] - \frac{\nu^2}{2}, \quad (36a)$$

$$t_n = \frac{(n+1)(n+\mu+\nu+1)(2n+\mu+\nu+\sigma+1)(2n+\mu+\nu-\sigma+1)}{2(2n+\mu+\nu+1)(2n+\mu+\nu+2)}, \quad (36b)$$

$$s_{n-1} = \frac{(n+\mu)(n+\nu)(2n+\mu+\nu+\sigma+1)(2n+\mu+\nu-\sigma+1)}{2(2n+\mu+\nu)(2n+\mu+\nu+1)}, \quad (36c)$$

The three-term recursion relation (6) with these coefficients and argument $z$ is identical to the recursion relation satisfied by a discrete version of the Wilson polynomial, which we designate as $\mathcal{W}_n(z)$ parametrized by $\mu$, $\nu$, and $\sigma$. The conventional Wilson polynomial, $W_n(z)$, forms an infinite sequence that has a mix of positive continuous spectrum and negative finite discrete spectrum [22]. Finally, the un-normalized *k*th bound state wavefunction (1) becomes

$$\psi_k(x) = [\sin(\rho x)]^{\mu+\frac{1}{2}} [\cos(\rho x)]^{-\mu-\nu-\frac{1}{2}} \sum_{n=0}^{N} \mathcal{W}_n\left(-E_k/\rho^2\right) J_n^{(\mu,\nu)}(y). \tag{37}$$

Now, the Pöschl-Teller potential (33) is exactly solvable, which gives us an opportunity to test our findings. The exact energy spectrum and corresponding wavefunctions are as follows [23]

$$E_k = 2\rho^2 \left(k + \frac{\mu+\sigma+1}{2}\right)^2, \tag{38}$$

$$\psi_k(x) = [\sin(\rho x)]^{\mu+\frac{1}{2}} [\cos(\rho x)]^{\sigma+\frac{1}{2}} P_k^{(\mu,\sigma)}(\cos(2\rho x)). \tag{39}$$

Figure 2 is a plot of these exact bound states wavefunction (as dotted blue curve) corresponding to the lowest part of the energy spectrum for a given set of physical parameters $\{\rho, V_0, V_1\}$. We superimpose on those the wavefunctions (37) (as solid red curve) for $\nu = -25$ that gives $N = \left\lfloor -\frac{\mu+\nu+1}{2} \right\rfloor = 10$. To demonstrate the convergence of our solution (37) with the size of the basis $N+1$, we plot in Figure 3 one of the exact wavefunctions and superimpose on it the wavefunction (37) for several values of *N*. It is evident that the matching with the exact result (i.e., convergence) improves with the basis size and that a very good match is obtained for small enough *N*.

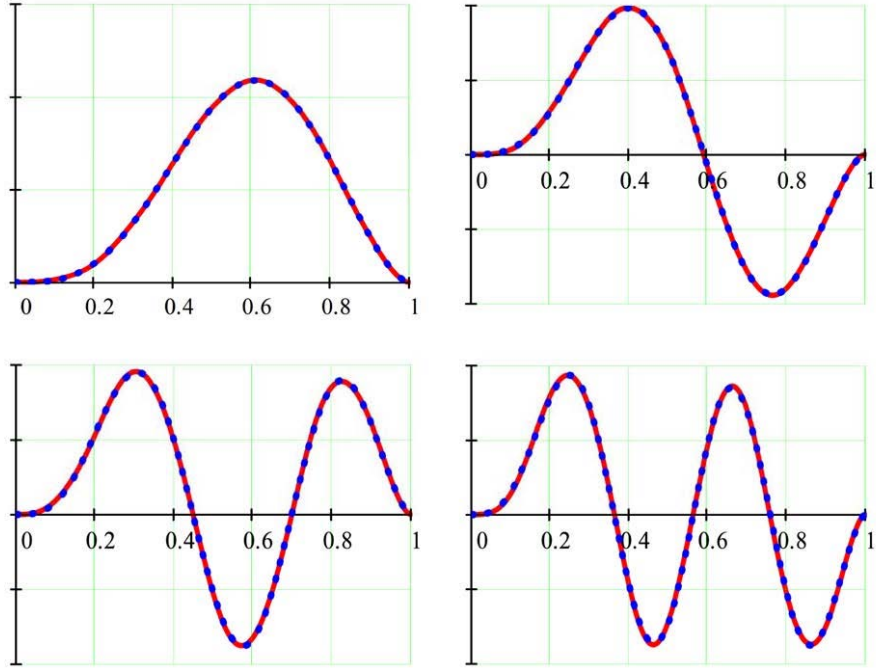

**Fig. 2:** The lowest energy eigenstates of the Pöschl-Teller potential as given by Eq. (37) (in red solid curve) for the potential parameters $\{\rho, V_0, V_1\} = \{1.0, 3.7, 0.5\}$ in atomic units. We super-imposed the exact bound states wavefunctions (39) as blue

dotted curves. The basis parameter was taken as $\nu = -25$, which gives $N = 10$. The horizontal *x*-axis is in units of $\pi/2\rho$.

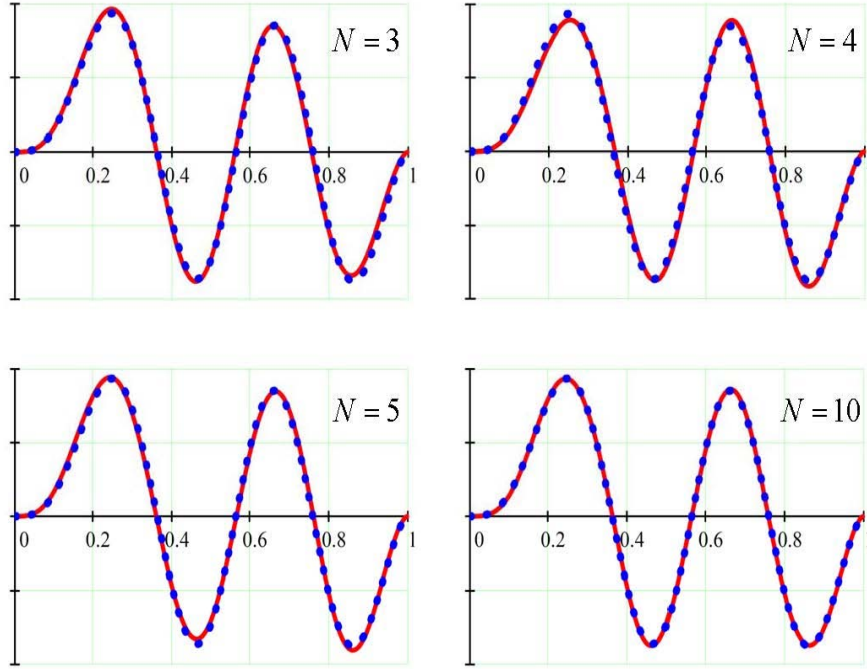

**Fig. 3:** Plots of the third excited state for the Pöschl-Teller potential as given by Eq. (37) and shown as a red solid curve with the basis parameter $\nu = -11, -13, -15, -25$ that corresponds to $N = 3, 4, 5, 10$, respectively. The blue dotted curve is the exact solution given by Eq. (39). Improvements in accuracy with the basis size is evident.

## 6. Conclusion

In this work, we demonstrated that a finite series of square integrable functions could give a very good approximation for bound states. As expected, the accuracy of such finite representation improves by increasing the number of these (basis) functions. Moreover, if the functions are chosen to give a tridiagonal matrix representation for the wave operator then we only need a small number of them for an accurate enough representation. We found two types of these functions written in terms of either the Romanovski-Bessel polynomials or the Romanovski-Jacobi polynomials.

The expansion coefficients of the series are finite orthogonal polynomials in the energy and/or potential parameters. They contain all physical information about the system. We gave several illustrative examples to support our findings and used the tridiagonal representation approach as a method of solution. We believe that this technique gives a physically faithful representation for systems with a finite number of bound states provided that the size of the spectrum is less than or equal to the size of the basis set.

–17–

## Appendix A: Romanovski-Bessel polynomials

For ease of reference, we reproduce the Appendix in our work [24] that gives all relevant properties of the R-Bessel polynomial on the real line.

The R-Bessel polynomial on the positive real line is defined in terms of the hypergeometric or confluent hypergeometric functions as follows (see section 9.13 of the book by Koekoek *et. al* [22] but make the replacement $x \mapsto 2x$ and $a \mapsto 2\mu$)

$$Y_n^\mu(x) = {}_2F_0\left(\begin{array}{c}-n, n+2\mu+1\\ -\end{array}\middle| -x\right) = (n+2\mu+1)_n x^n \, {}_1F_1\left(\begin{array}{c}-n\\ -2(n+\mu)\end{array}\middle| 1/x\right), \tag{A1}$$

where $x \geq 0$, $n = 0, 1, 2, ..., N$ and $N$ is a non-negative integer. The real parameter $\mu$ is negative such that $\mu < -N - \tfrac{1}{2}$. The Pochhammer symbol $(a)_n$ (a.k.a. shifted factorial) is defined as $(a)_n = a(a+1)(a+2)...(a+n-1) = \frac{\Gamma(n+a)}{\Gamma(a)}$. The R-Bessel polynomial could also be written in terms of the associated Laguerre polynomial as: $Y_n^\mu(x) = n!(-x)^n L_n^{-(2n+2\mu+1)}(1/x)$. The three-term recursion relation reads as follows:

$$2x Y_n^\mu(x) = \frac{-\mu}{(n+\mu)(n+\mu+1)} Y_n^\mu(x)$$
$$- \frac{n}{(n+\mu)(2n+2\mu+1)} Y_{n-1}^\mu(x) + \frac{n+2\mu+1}{(n+\mu+1)(2n+2\mu+1)} Y_{n+1}^\mu(x) \tag{A2}$$

Note that the constraints on $\mu$ and on the maximum polynomial degree make this recursion definite (i.e., the signs of the two recursion coefficients multiplying $Y_{n\pm 1}^\mu(x)$ are the same). Otherwise, these polynomials could not be defined on the real line but on the unit circle in the complex plane. The orthogonality relation reads as follows

$$\int_0^\infty x^{2\mu} e^{-1/x} Y_n^\mu(x) Y_m^\mu(x) dx = -\frac{n! \Gamma(-n-2\mu)}{2n+2\mu+1} \delta_{nm}. \tag{A3}$$

The differential equation is

$$\left\{x^2 \frac{d^2}{dx^2} + [1 + 2x(\mu+1)] \frac{d}{dx} - n(n+2\mu+1)\right\} Y_n^\mu(x) = 0. \tag{A4}$$

The forward and backward shift differential relations read as follows

$$\frac{d}{dx} Y_n^\mu(x) = n(n+2\mu+1) Y_{n-1}^{\mu+1}(x). \tag{A5}$$

$$x^2 \frac{d}{dx} Y_n^\mu(x) = -(2\mu x + 1) Y_n^\mu(x) + Y_{n+1}^{\mu-1}(x). \tag{A6}$$

We can write $Y_{n+1}^{\mu-1}(x)$ in terms of $Y_n^\mu(x)$ and $Y_{n\pm 1}^\mu(x)$ as follows

$$2Y_{n+1}^{\mu-1}(x) = \frac{(n+1)(n+2\mu)}{(n+\mu)(n+\mu+1)} Y_n^\mu(x)$$
$$+ \frac{n(n+1)}{(n+\mu)(2n+2\mu+1)} Y_{n-1}^\mu(x) + \frac{(n+2\mu)(n+2\mu+1)}{(n+\mu+1)(2n+2\mu+1)} Y_{n+1}^\mu(x) \tag{A7}$$

Using this identity and the recursion relation (A2), we can rewrite the backward shift differential relation as follows

$$2x^2 \frac{d}{dx} Y_n^\mu(x) = n(n+2\mu+1) \times$$
$$\left[ -\frac{Y_n^\mu(x)}{(n+\mu)(n+\mu+1)} + \frac{Y_{n-1}^\mu(x)}{(n+\mu)(2n+2\mu+1)} + \frac{Y_{n+1}^\mu(x)}{(n+\mu+1)(2n+2\mu+1)} \right] \tag{A8}$$

The generating function is

$$\sum_{n=0}^\infty Y_n^\mu(x) \frac{t^n}{n!} = \frac{2^{2\mu}}{\sqrt{1-4xt}} \left(1+\sqrt{1-4xt}\right)^{-2\mu} \exp\left[2t/(1+\sqrt{1-4xt})\right]. \tag{A9}$$

The polynomial $B_n^\mu(z;\gamma)$ is defined in [25] by its three-term recursion relation Eq. (15) therein, which reads

$$z B_n^\mu(z;\gamma) = \left[ \frac{-2\mu}{(n+\mu)(n+\mu+1)} + \gamma \left(n+\mu+\tfrac{1}{2}\right)^2 \right] B_n^\mu(z;\gamma)$$
$$- \frac{n}{(n+\mu)\left(n+\mu+\tfrac{1}{2}\right)} B_{n-1}^\mu(z;\gamma) + \frac{n+2\mu+1}{(n+\mu+1)\left(n+\mu+\tfrac{1}{2}\right)} B_{n+1}^\mu(z;\gamma) \tag{A10}$$

where $B_0^\mu(z;\gamma) = 1$ and $B_{-1}^\mu(z;\gamma) := 0$.

## Appendix B: Romanovski-Jacobi polynomials

For ease of reference, we reproduce Appendix A in our work [26] that gives all relevant properties of these finite R-Jacobi polynomials.

These polynomials are defined on the semi-infinite interval $y \geq 1$ whereas the conventional Jacobi polynomials $P_n^{(\mu,\nu)}(y)$ are defined on the finite interval $-1 \leq y \leq +1$. To distinguish it from the polynomials $P_n^{(\mu,\nu)}(y)$ we use the notation $J_n^{(\mu,\nu)}(y)$ for the R-Jacobi polynomial:

$$J_n^{(\mu,\nu)}(y) = \frac{\Gamma(n+\mu+1)}{\Gamma(n+1)\Gamma(\mu+1)} {}_2F_1\left( \begin{matrix} -n, n+\mu+\nu+1 \\ \mu+1 \end{matrix} \middle| \frac{1-y}{2} \right) = (-1)^n J_n^{(\nu,\mu)}(-y). \tag{B1}$$

where $n = 0, 1, 2, ..., N$, $\mu > -1$ and $\mu + \nu < -2N - 1$. It satisfies the following differential equation

$$\left\{ (y^2 - 1) \frac{d^2}{dy^2} + \left[ (\mu+\nu+2)y + \mu - \nu \right] \frac{d}{dy} - n(n+\mu+\nu+1) \right\} J_n^{(\mu,\nu)}(y) = 0, \tag{B2}$$

It also satisfies the following three-term recursion relation

$$y J_n^{(\mu,\nu)}(y) = \frac{\nu^2 - \mu^2}{(2n+\mu+\nu)(2n+\mu+\nu+2)} J_n^{(\mu,\nu)}(y)$$
$$+ \frac{2(n+\mu)(n+\nu)}{(2n+\mu+\nu)(2n+\mu+\nu+1)} J_{n-1}^{(\mu,\nu)}(y) + \frac{2(n+1)(n+\mu+\nu+1)}{(2n+\mu+\nu+1)(2n+\mu+\nu+2)} J_{n+1}^{(\mu,\nu)}(y) \tag{B3}$$

and the following differential relation

$$(y^2 - 1) \frac{d}{dy} J_n^{(\mu,\nu)} = 2(n+\mu+\nu+1) \left[ \frac{(\nu-\mu)n}{(2n+\mu+\nu)(2n+\mu+\nu+2)} J_n^{(\mu,\nu)} \right.$$
$$\left. - \frac{(n+\mu)(n+\nu)}{(2n+\mu+\nu)(2n+\mu+\nu+1)} J_{n-1}^{(\mu,\nu)} + \frac{n(n+1)}{(2n+\mu+\nu+1)(2n+\mu+\nu+2)} J_{n+1}^{(\mu,\nu)} \right] \tag{B4}$$

The associated orthogonality relation for the R-Jacobi polynomials reads as follows

$$\int_1^\infty (y-1)^\mu (y+1)^\nu J_n^{(\mu,\nu)}(y) J_m^{(\mu,\nu)}(y) dy = \frac{2^{\mu+\nu+1}}{2n+\mu+\nu+1} \frac{\Gamma(n+\mu+1)\Gamma(n+\nu+1)}{\Gamma(n+1)\Gamma(n+\mu+\nu+1)} \frac{\sin \pi \nu}{\sin \pi(\mu+\nu+1)} \delta_{nm}, \tag{B5}$$

where $n, m \in \{0, 1, 2, ..., N\}$. Equivalently (see Eq. 4.9 in Ref. [27]),

$$\int_1^\infty (y-1)^\mu (y+1)^\nu J_n^{(\mu,\nu)}(y) J_m^{(\mu,\nu)}(y) dy = \frac{(-1)^{n+1} 2^{\mu+\nu+1}}{2n+\mu+\nu+1} \frac{\Gamma(n+\mu+1)\Gamma(n+\nu+1)\Gamma(-n-\mu-\nu)}{\Gamma(n+1)\Gamma(-\nu)\Gamma(\nu+1)} \delta_{nm}. \tag{B6}$$

The polynomial $H_n^{(\mu,\nu)}(\zeta;\gamma,\theta)$ is defined in Ref. [19] by its three-term recursion relation, which is Eq. (8) therein that reads

$$(\cos\theta) H_n^{(\mu,\nu)}(\zeta;\gamma,\theta) = \left\{ \left[ \left(n + \frac{\mu+\nu+1}{2}\right)^2 - \gamma^2 \right] \zeta \sin\theta + \frac{\nu^2 - \mu^2}{(2n+\mu+\nu)(2n+\mu+\nu+2)} \right\} H_n^{(\mu,\nu)}(\zeta;\gamma,\theta)$$
$$+ \frac{2(n+\mu)(n+\nu)}{(2n+\mu+\nu)(2n+\mu+\nu+1)} H_{n-1}^{(\mu,\nu)}(\zeta;\gamma,\theta) + \frac{2(n+1)(n+\mu+\nu+1)}{(2n+\mu+\nu+1)(2n+\mu+\nu+2)} H_{n+1}^{(\mu,\nu)}(\zeta;\gamma,\theta), \tag{B7}$$

where $0 > \theta > \pi$, $H_0^{(\mu,\nu)}(\zeta;\gamma,\theta) = 1$ and $H_{-1}^{(\mu,\nu)}(\zeta;\gamma,\theta) := 0$. For some range of values of the polynomial parameters, it is more appropriate to define $\tilde{H}_n^{(\mu,\nu)}(\zeta;\gamma,\theta) = H_n^{(\mu,\nu)}(-i\zeta;\gamma,i\theta)$, which maps the recursion (B7) into

$$(\cosh\theta) \tilde{H}_n^{(\mu,\nu)}(\zeta;\gamma,\theta) = \left\{ \left[ \left(n + \frac{\mu+\nu+1}{2}\right)^2 - \gamma^2 \right] \zeta \sinh\theta + \frac{\nu^2 - \mu^2}{(2n+\mu+\nu)(2n+\mu+\nu+2)} \right\} \tilde{H}_n^{(\mu,\nu)}(\zeta;\gamma,\theta)$$
$$+ \frac{2(n+\mu)(n+\nu)}{(2n+\mu+\nu)(2n+\mu+\nu+1)} \tilde{H}_{n-1}^{(\mu,\nu)}(\zeta;\gamma,\theta) + \frac{2(n+1)(n+\mu+\nu+1)}{(2n+\mu+\nu+1)(2n+\mu+\nu+2)} \tilde{H}_{n+1}^{(\mu,\nu)}(\zeta;\gamma,\theta), \tag{B8}$$